\documentclass[%
 reprint,
superscriptaddress,
 amsmath,amssymb,
 prl,
]{revtex4-2}

\usepackage{titlesec}%
\usepackage{subfigure}%
\usepackage{amsfonts}%
\usepackage{graphicx}%
\usepackage{setspace}%
\usepackage{comment}%
\usepackage{dsfont}%
\usepackage{color}%
\usepackage{tikz}
\usetikzlibrary{arrows,calc}
\usepackage[pdftex,colorlinks=true,linkcolor=blue,citecolor=blue]{hyperref}

\newcommand{\Rv}{{\bf R}}
\newcommand{\ii}{\mathrm{i}}
\newcommand{\subl}{\mathcal{S}}%
\newcommand{\tr}{\mathrm{Tr}} 

\newcommand{\ket}[1]{\left| #1 \right>} 
\newcommand{\bra}[1]{\left< #1 \right|} 

\titleformat{\paragraph}
{\normalfont\normalsize\bfseries}{\theparagraph}{1em}{}
\titlespacing*{\paragraph}
{0pt}{3.25ex plus 1ex minus .2ex}{1.5ex plus .2ex}

\begin{document}

\title{Chiral-Symmetric Higher-Order Topological Phases of Matter}

\author{Wladimir A. Benalcazar}
\email{wb7707@princeton.edu}
\affiliation{Department of Physics, Princeton University, Princeton, New Jersey 08542, USA}
\affiliation{Department of Physics, Pennsylvania State University, University Park, Pennsylvania 16802, USA}

\author{Alexander Cerjan}
\affiliation{Center for Integrated Nanotechnologies, Sandia National Laboratories, Albuquerque, New Mexico 87123, USA}
\date{\today}

\begin{abstract}
We introduce novel higher-order topological phases of matter in chiral-symmetric systems (class AIII of the tenfold classification), most of which would be misidentified as trivial by current theories. These phases are protected by ``multipole chiral numbers", bulk integer topological invariants that in 2D and 3D are built from sublattice multipole moment operators, as defined herein. The integer value of a multipole chiral number indicates how many degenerate zero-energy states localize at each corner of a system. These higher-order topological phases of matter are generally boundary-obstructed and robust in the presence of chiral-symmetry-preserving disorder.
\end{abstract}

\maketitle

Higher-order topological band theory has expanded the classification of topological phases of matter across insulators~\cite{benalcazar2017quad,song2017,langbehn2017,benalcazar2017quadPRB,ortix2018,schindler2018hoti,khalaf2018hoti,geier2018,benalcazar2019fillinganomaly,yue2019,xu2019,elcoro2021,xu2020_high-throughput}, semimetals~\cite{Lin17,Ahn18,wang2019,wieder2020,ghorashi2020,xu2020_high-throughput}, and superconductors~\cite{benalcazar2014,wang2018,liu2018,dwivedi2018,zhu2018,hsu2018,ghorashi2019vortex,franca2019,yan2019,vol2019,zhang2020kitaev,vu2020,schindler2020,li2021higherorder,scammell2021intrinsic}. It generalizes the bulk-boundary correspondence of topological phases, so that an $n$th-order topological phase in $d$ dimensions has protected features, such as gapless states or fractional charges, only at its $(d-n)$-dimensional boundaries. Currently, two complementary mechanisms are known to give rise to higher-order topological phases (HOTPs): (i) corner-induced filling anomalies due to certain Wannier center configurations~\cite{song2017,ortix2018,benalcazar2019fillinganomaly,fang2021,takahashi2021}, and (ii) the existence of boundary-localized mass domains~\cite{song2017,langbehn2017,schindler2018hoti,geier2018,khalaf2018,khalaf2018hoti,trifunovic2019}. These two mechanisms protect the fractional quantization of corner charge and the existence of single in-gap states at corners, respectively.
 
In first-order topological systems, phases protecting multiple states at each boundary also exist. This occurs in chiral-symmetric systems (class AIII in the tenfold classification~\cite{schnyder2008,kitaev2009,ryu2010}) in odd dimensions. In 1D, for example, such phases are identified by a $\mathbb{Z}$ topological invariant known as the winding number~\cite{ryu2002,teo2010}, which classifies the Hamiltonian's homotopy class within the first homotopy group $\pi_1[U(N)]$ and indicates the number of degenerate zero-energy states at each boundary. 
In contrast, the Wannier center approach applied to chiral 1D systems only yields a $\mathbb{Z}_2$ classification according to whether the electric dipole moment (given by the position of the Wannier centers) is quantized to 0 or $e/2$. In particular, it labels all 1D chiral-symmetric systems with even winding numbers as trivial.

The observation that 1D systems in class AIII have a broader classification than the one provided by the Wannier center picture suggests that, analogously, a broader classification could exist for HOTPs in class AIII. Consider, for example, stacking $N$ topological quadrupole insulators~\cite{benalcazar2017quad}. If they are coupled in a chiral-symmetric fashion, the overall system will have $N$ zero-energy states protected at each corner. Yet, the topological invariants that protect this phase have not been found. Moreover, the existence of such broader classification would apparently be at odds with the tenfold classification of topological phases, which predicts only trivial phases for chiral-symmetric systems in 2D. This prediction stems from the fact that higher-dimensional generalizations of the 1D winding number ---which identify classes within the homotopy group $\pi_{d}[U(N)]$ in $d$ dimensional systems--- are trivial for even $d$ 
~\cite{nakahara}. The resolution to this apparent contradiction is that the tenfold classification applies to first-order, bulk-obstructed topological phases, while the phases we consider here are higher order and boundary obstructed. Hence, a different approach is needed to classify chiral-symmetric HOTPs, i.e., one that goes beyond the natural generalization of the 1D winding number to higher dimensions.

In this Letter, we demonstrate the existence of a $\mathbb{Z}$ classification for HOTPs in class AIII and identify the topological invariants in 2D and 3D that protect them.
We refer to these invariants as multipole chiral numbers (MCNs) because they  generalize the classification provided by the 1D winding number to higher-dimensional systems but, instead of being the traditional generalization of winding numbers to higher dimensions~\cite{teo2010}, they are built from sublattice multipole moment operators, and capture higher-order, boundary-obstructed topology~\cite{benalcazar2017quadPRB,khalaf2021,ezawa2020botp,tiwari2020botp,asaga2020botp,wu2020botp,tiwari2020botp}.
These invariants are calculated in the bulk of the system, i.e., with periodic boundary conditions, and, for sufficiently large systems, their integer values indicate the number of degenerate zero-energy states at each corner of a system with open boundaries. Thus, MCNs provide a higher-order bulk-boundary correspondence for topological phases in class AIII. Since MCNs are defined in real space, they can be used to characterize disordered systems, and here we demonstrate that phases protected by MCNs are robust in the presence of chiral-symmetry-preserving disorder.
The existence of phases with MCNs reveals a richer classification of HOTPs, provides a broader understanding of boundary-obstructed topological phases beyond the Wannier center and mass domain perspectives, and has implications for the further classification of HOTPs in interacting systems~\cite{fidkowski2010}. The phases we present can be readily proven in several synthetic material platforms~\cite{peterson2018,kollar2019,imhof2018,deng2021}, and recent advances on the generation and control of long-range hoppings could enable their realization in ultracold atoms in optical lattices~\cite{aidelsburger2013,miyake2013,jotzu2014,martinez2021}.

Let us focus our attention on chiral-symmetric Hamiltonians ${\cal H}$, which satisfy
$\Pi {\cal H} \Pi = -{\cal H}$, where $\Pi$ is the chiral operator.
In the basis in which the chiral operator is $\Pi=\tau_z$, the Hamiltonian ${\cal H}$ takes the form
\begin{align}
{\cal H}=\begin{pmatrix}
0 & h\\
h^\dagger & 0
\end{pmatrix},
\label{eq:ChiralHamiltonian}
\end{align}
which allows a partition of the lattice into two sublattices, $A$ and $B$, with opposite chiral charge. 
The eigenstates of ${\cal H}$ can be written as $\ket{\psi_n}=\frac{1}{\sqrt{2}}(\psi^A_n, \psi^B_n)^T$, where $\psi^A_n$ and $\psi^B_n$ are normalized vectors that exist only in the $A$ and $B$ subspaces, respectively. 
Under chiral symmetry, every eigenstate $\ket{\psi_n}$ with energy $\epsilon_n$ has a chiral partner state $\Pi \ket{\psi_n}=\frac{1}{\sqrt{2}}(\psi^A_n, -\psi^B_n)^T$ with energy $-\epsilon_n$. Evaluating $\mathcal{H}^2\ket{\psi_n}=\epsilon_n^2 \ket{\psi_n}$ leads to the eigenvalue problems $(h h^\dagger) \psi^A_n = \epsilon_n^2 \psi^A_n$ and $(h^\dagger h) \psi^B_n = \epsilon_n^2 \psi^B_n$,
so that $\psi^A_n$ and $\psi^B_n$ can be easily obtained by diagonalizing $hh^\dagger$ or $h^\dagger h$, respectively. This structure is reflected in the singular value decomposition (SVD) of $h$ by writing 
\begin{align}
h=U_A \Sigma U_B^\dagger,
\label{eq:SVD}
\end{align}
where $U_\subl$, for $\subl=A,B$, is a unitary matrix representing the space spanned by $\{\psi^\subl_n\}$, i.e., $U_\subl=(\psi^\subl_1, \psi^\subl_2 \ldots, \psi^\subl_{N_\subl})$, and $\Sigma$ is a diagonal matrix containing the singular values. Using this decomposition, it follows that $hh^\dagger = U_A \Sigma^2 U_A^\dagger$ and $h^\dagger h = U_B \Sigma^2 U_B^\dagger$, so that the squared energies $\{\epsilon^2_n\}$ correspond to the squared singular values in $\Sigma^2$. 

The SVD decomposition~\eqref{eq:SVD} allows an explicit flattening of the Hamiltonian by defining the unitary matrix $q=U_A U_B^\dagger$. The winding number of a Bloch Hamiltonian in 1D parametrized by the crystal momentum $k$ is then given by $N_x=(1/2\pi \ii)\int_{BZ} \tr \left[ q(k)^\dagger \partial_k q(k) \right]$, and is a topological invariant associated with the homotopy classes in $\pi_1[U(n)]=\mathbb{Z}$.

In the absence of periodicity, $k$ is not a good quantum number and the winding number loses its meaning. However, it is still possible to find real space topological invariants of chiral-symmetric 1D systems (equivalent to the winding number when periodicity is restored), which have allowed for the study of the effects of disorder~\cite{mondragon2014,lin2021realspace,velury2021}. Specifically, the 1D winding number is equivalent to the real space index
$N_x =(1/2\pi \ii) \tr \mathrm{log}(\bar{P}_x^A \bar{P}_x^{B\dagger}) \in \mathbb{Z}$, where $\bar{P}_x^\subl = U_\subl^\dagger P^\subl_x U_\subl$ is the sublattice dipole operator projected into the spaces $U_\subl$, for $\subl=A, B$~\cite{SuppInfo,lin2021realspace}. 
Here, $P_x^\subl$ is defined using the dipole moment operator for periodic systems~\cite{resta1998}, but restricted to a single sublattice, 
$P^\subl_x=\sum_{R,\alpha \in \subl} \ket{R,\alpha} \text{Exp}(-\ii 2\pi R/L) \bra{R,\alpha}$,
where the 1D crystal has $L$ unit cells, $\ket{R,\alpha}=c^\dagger_{R,\alpha}\ket{0}$, and $c^\dagger_{R,\alpha}$ creates an electron at orbital $\alpha$ of unit cell $R$.

The MCNs for higher-order topological phases with chiral symmetry are based on extensions of this formulation of real space indices to 2D and 3D. 
Consider a lattice in 2D (3D) with $L_j$ unit cells along direction $j=x,y$ ($j=x,y,z$). Each unit cell is labeled by $\Rv=(x,y)$ [$\Rv=(x,y,z)$] and has $N_T$ orbitals (or, more generally, $N_T$ internal degrees or freedom). 
To build the topological indices for chiral-symmetric higher-order topological phases,
we define the following sublattice multipole moment operators:
\begin{align}
Q^\subl_{xy}&=\sum_{\Rv,\alpha \in \subl} \ket{\Rv,\alpha}\text{Exp}\left(-\ii \frac{2\pi xy}{L_x L_y}\right) \bra{\Rv,\alpha}\label{eq:Operators1}
\\
O^\subl_{xyz}&=\sum_{\Rv,\alpha \in \subl} \ket{\Rv,\alpha}\text{Exp}\left(-\ii \frac{2\pi xyz}{L_x L_y L_z}\right) \bra{\Rv,\alpha},
\label{eq:Operators2}
\end{align}
for 2D and 3D lattices, respectively. These operators resemble those associated with quadrupole and octupole moments~\cite{wheeler2019,kang2019,watanabe2019quad}, but are only defined over each sublattice $\subl=A,B$, instead of across the entire system.

We claim that the integer invariants for chiral-symmetric second-order topological phases in 2D and third-order topological phases in 3D are, respectively,
\begin{align}
N_{xy}&=\frac{1}{2\pi \ii} \tr \mathrm{log} \left(\bar{Q}_{xy}^A \bar{Q}_{xy}^{B\dagger} \right) \in \mathbb{Z}
\label{eq:Invariants1}\\
N_{xyz}&=\frac{1}{2\pi \ii} \tr \mathrm{log} \left(\bar{O}_{xyz}^A \bar{O}_{xyz}^{B\dagger} \right)\in \mathbb{Z},
\label{eq:Invariants2}
\end{align}
where $\bar{Q}_{xy}^\subl=U_\subl^\dagger Q^\subl_{xy} U_\subl$ and $\bar{O}^\subl_{xyz}=U_\subl^\dagger O^\subl_{xyz} U_\subl$, for $\subl=A,B$, are the sublattice multipole moment operators projected into the spaces $U_\subl$.
To demonstrate that Eqs.\ \eqref{eq:Invariants1} and \eqref{eq:Invariants2} are the invariants for chiral-symmetric higher-order topological phases, one must show that these invariants are strictly quantized, that they predict the number of topologically protected corner states at each corner of the lattice, and that phases with different MCNs are separated from one another by phase transitions that close the energy gap.

To prove that the invariants~\eqref{eq:Invariants1} and \eqref{eq:Invariants2} are strictly quantized, notice that they take the form $N=(1/2\pi \ii) \tr \mathrm{log} (U_A^\dagger M_A U_A U_B^\dagger M_B^\dagger U_B)$, where $M_\subl$ (for $\subl=A,B$) is $Q_{xy}^\subl$ in 2D, or $O_{xyz}^\subl$ in 3D. Since the matrices $M_\subl$ and $U_\subl$ are unitary, we have $\det (U_A^\dagger M_A U_A U_B^\dagger M_B^\dagger U_B)= \det (M_A M^\dagger_B)=1$, where the last step follows if the two sublattices have (i) equal number of degrees of freedom in each unit cell and (ii) the same number of unit cells. Under these conditions, tracing the logarithm of $U_A^\dagger M_A U_A U_B^\dagger M^\dagger_B U_B$ will necessarily give a phase that is a multiple of $2\pi \ii$, i.e., it will be of the form $2\pi \ii N$, with $N \in \mathbb{Z}$. This integer $N$ is the topological invariant.
Exploiting this structure of the invariants, Eqs.~\eqref{eq:Invariants1} and \eqref{eq:Invariants2} can also be written in the form of a Bott index \cite{exel_invariants_1991,hastings_almost_2010}, see Supplemental Material~\cite{SuppInfo}.

\begin{figure}[t!]
\centering
\includegraphics[width=1.0\columnwidth]{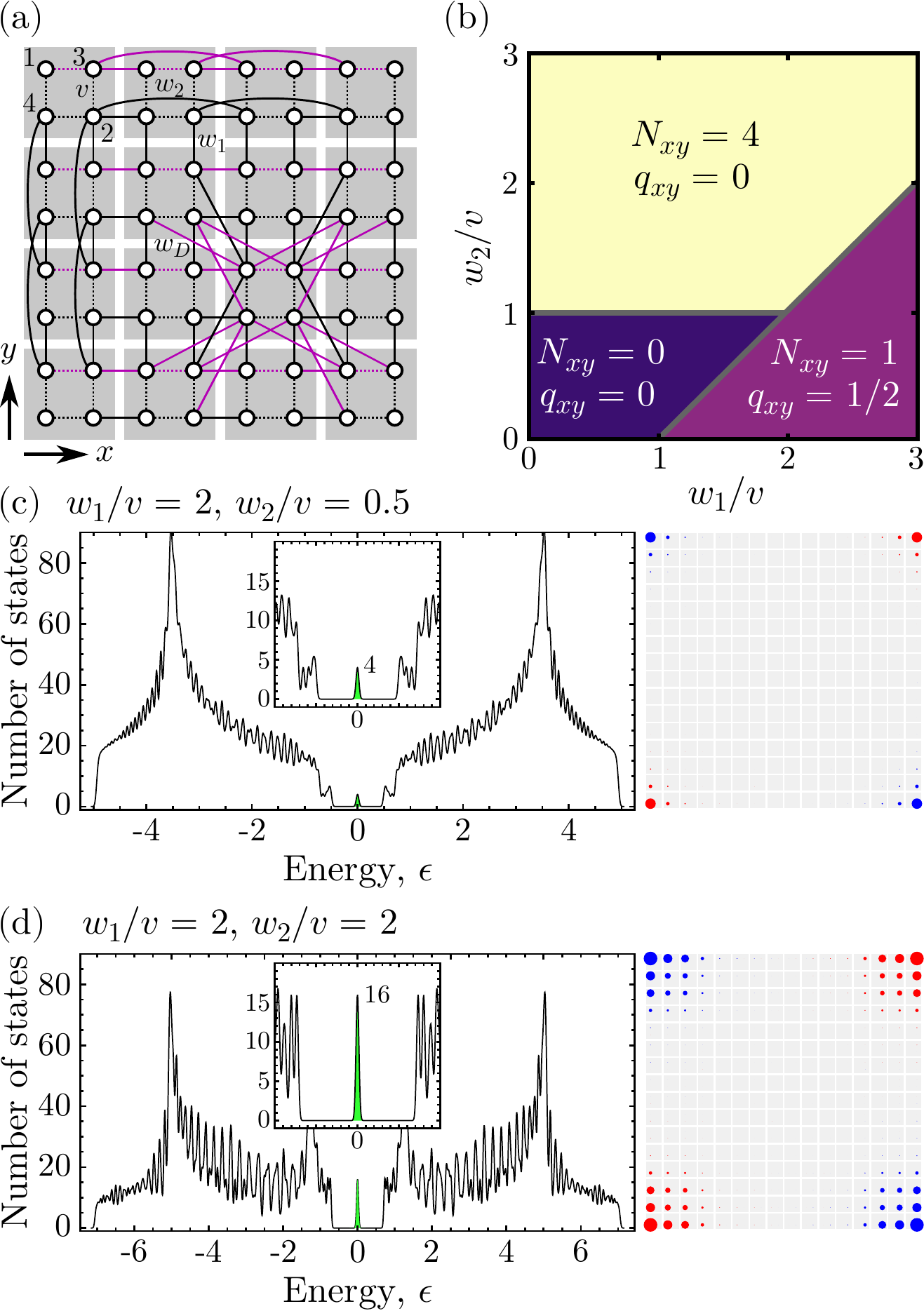}
\caption{(a) Schematic depicting the tight-binding model used. Not all non-nearest-neighbor hoppings are shown for clarity. All purple hoppings are multiplied by $-1$ such that each plaquette has a uniform flux of $\pi$. (b) Phase diagram indicating the quadrupole chiral number $N_{xy}$ and the quadrupole moment $q_{xy}$ for a $C_{4v}$-symmetric system. Here, $w_{m>2} = 0$ and $w_D = 0$. Different phases are separated by gray lines of critical points where the bulk band gap closes. (c,d) Density of states (left) and local density of states at zero energy (right) for the $N_{xy}=1$ phase (c) and the $N_{xy}=4$ phase (d). On the right sides of (c) and (d), red and blue colors indicate support over the $A$ and $B$ sublattices, respectively.}
\label{fig:1}
\end{figure}

We now illustrate some of the topological phases with nonzero values of $N_{xy}$ and demonstrate that this invariant corresponds to the number of corner-localized states in each corner. Consider the quadrupole topological insulator (QTI) \cite{benalcazar2017quad} with additional long-range hopping terms.
The Bloch Hamiltonian for the QTI has the form of Eq.\ (\ref{eq:ChiralHamiltonian}) with the off-diagonal matrix
\begin{align}
    h_{\textrm{QTI}}(\mathbf{k}) =& \left( \begin{array}{cc}
    -v_x -w_{1,x} e^{-\ii k_x} & v_y + w_{1,y} e^{\ii k_y} \\
    v_y + w_{1,y} e^{-\ii k_y} & v_x + w_{1,x} e^{\ii k_x} \end{array} \right), \label{eq:2dssh}
\end{align}
where $v_{x/y}$ and $w_{1,x/y}$ are the nearest-neighbor hoppings within a unit cell and between adjacent unit cells, respectively (generally, we allow for different values of these hoppings in the $x$ and $y$ directions). Adding to this model, we also allow for straight long-range (SLR) hoppings,
\begin{align}
    h_{\textrm{SLR}}(\mathbf{k}) =& \sum_{m>1}^M \left( \begin{array}{cc}
    -w_{m,x} e^{-\ii m k_x} & w_{m,y} e^{\ii m k_y} \\
    w_{m,y} e^{-\ii m k_y} & w_{m,x} e^{\ii m k_x} \end{array} \right), \label{eq:slr}
\end{align}
where $M$ determines the maximum long-range hopping, as well as diagonal long-range (DLR) hoppings,
\begin{align}
    h_{\textrm{DLR}}(\mathbf{k}) = 2 w_D \left( \begin{array}{cc}
    e^{-\ii k_x} \cos(k_y) & -e^{\ii k_y} \cos(k_x) \\
    -e^{-\ii k_y} \cos(k_x) & -e^{\ii k_x} \cos(k_y) \end{array} \right). \label{eq:chiralDiag}
\end{align}
Here, $w_{m,x/y}$ are the long-range hoppings among the $m$th nearest-neighbor unit cells in the horizontal and vertical direction, and $w_D$ are hoppings among nearest-neighbor unit cells along the diagonal directions. All the terms preserve chiral symmetry and the diagonal terms \eqref{eq:chiralDiag} break separability, making it impossible to write the full Hamiltonian as $\mathcal{H}(\mathbf{k}) = \mathcal{H}_x(k_x) + \mathcal{H}_y(k_y)$.
In writing this Hamiltonian, we thread a $\pi$ flux through each plaquette of the system, which is implemented via the specific choice of gauge directly written in Eqs.\ (\ref{eq:2dssh}-\ref{eq:chiralDiag}) and shown in Fig.~\ref{fig:1}(a).

First, consider a chiral and $C_{4}$-symmetric, long-range QTI model with $w_{m>2}=0$ and $w_D = 0$. For $w_{m}/v < 1$, this system possesses a bulk band gap around zero energy and both the quadrupole moment $q_{xy}$~\cite{benalcazar2017quad} and the quadrupole winding number $N_{xy}$ [Eq.~\ref{eq:Invariants1}] identify it as trivial ($q_{xy}=0$, $N_{xy}=0$), Fig.~\ref{fig:1}(b). Starting from this phase and increasing $w_1/v$, a bulk band-gap-closing phase transition occurs, after which both topological indices now show that this system is in a nontrivial phase ($q_{xy}=1/2$, $N_{xy}=1$). With open boundaries, this phase possesses a single zero-energy state localized to each of its corners, Fig.~\ref{fig:1}c. This is the previously known QTI phase~\cite{benalcazar2017quad}. However, when the long-range hopping $w_2/v$ is increased, a separate bulk band gap-closing phase transition occurs that  separates either the $N_{xy}=0$ phase or the $N_{xy} = 1$ phase from another nontrivial phase with $N_{xy} = 4$, but with $q_{xy} = 0$. Simulations of the open system reveal that each corner of the lattice in this phase possesses four degenerate states with $\epsilon=0$ and that all such states within a corner exist only on a single sublattice of the system, see Fig.~\ref{fig:1}(d) and Fig.~S2 in the Supplemental Material~\cite{SuppInfo}. 

Since all of the zero-energy states within a corner occupy the same sublattice, they have the same chiral charge $\Pi |\psi_{\textrm{corner}} \rangle = \pm | \psi_{\textrm{corner}} \rangle$ and, thus, cannot pair to hybridize away from zero energy as long as chiral symmetry is preserved.

Not only is the $N_{xy}=4$ phase not captured by the quadrupole index, but more generally, it lies beyond the framework of induced band representations~\cite{bradlyn2017,cano2018}. Consequently, topological indices based on calculating the representations of the bulk bands at high-symmetry points of the Brillouin zone will fail to find this phase, as the representations of the lowest two bands at all of the high-symmetry points are identical in the $N_{xy}=4$ phase, leading to trivial symmetry indicator invariants, see Supplemental Material~\cite{SuppInfo}.

\begin{figure}[t]
\centering
\includegraphics[width=1.0\columnwidth]{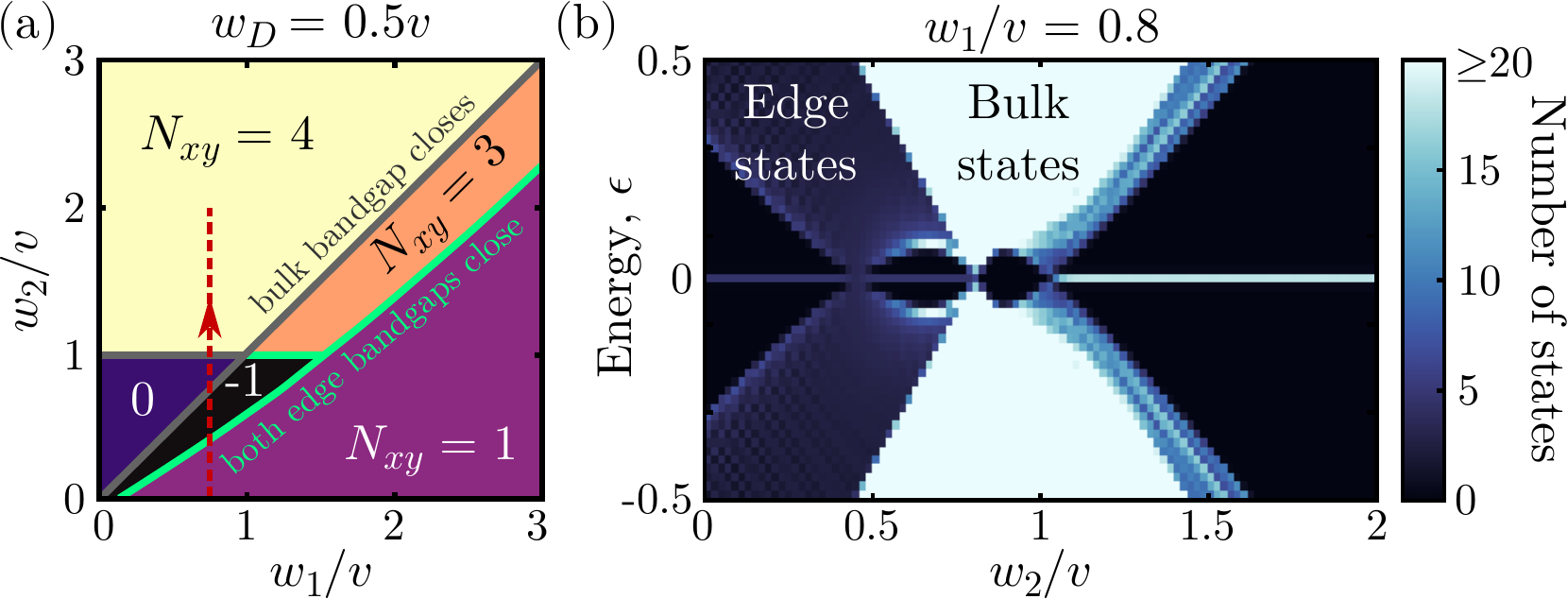}
\caption{(a) Phase diagram of the $N_{xy}$ phases for a $C_{4v}$-symmetric, separability-broken system with $w_D/v = 0.5$ and $w_{m>2} = 0$. Bulk-obstructed phase transitions are shown in gray, while boundary-obstructed phase transitions are shown in lime. (b) Density of states for this system for fixed $w_1/v = 0.8$, indicated as the red line in (a).}
\label{fig:noc4}
\end{figure}

Phase transitions between phases with different MCNs need not close the bulk band gap but, at a minimum, must close some lower-dimensional edge or surface band gap. HOTPs with this property are known as boundary-obstructed topological phases~\cite{khalaf2021}. This property remains true even in the presence of $C_4$ symmetry, which renders the QTI phase bulk obstructed. For example, consider adding diagonal long-range hoppings to this model, $w_D/v = 0.5$ [Eq.~\ref{eq:chiralDiag}], which preserve chiral and $C_4$ symmetries but break separability. As can be seen in Fig.~\ref{fig:noc4}(a), the $N_{xy} = -1$ and $N_{xy}=3$ phases each have a phase boundary in which the bulk band gap closes, and boundaries with other phases where only the edge band gap closes. Both of these types of boundaries can be explicitly seen in the density of states across these phase transitions, Fig.~\ref{fig:noc4}(b). For all of the different phases identified in Fig.~\ref{fig:noc4}(a), the number of states localized in each corner of the system is equal to $|N_{xy}|$ and the sublattice over which the corner states are supported is given by $\text{sgn}(N_{xy})$. Thus, for example, the $N_{xy} = -1$ phase in Fig.~\ref{fig:noc4}(a) indicates that the system possesses one state localized in each corner with support only on the \emph{opposite} sublattice when compared with those in phases with $N_{xy} > 0$, see Supplemental Material~\cite{SuppInfo}. In 3D, chiral-symmetric higher-order phases are characterized by distinct integer values of Eq.~\eqref{eq:Invariants2}, which indicate the number of degenerate states localized at each corner in the 3D structure. 

\begin{figure}[t]
\centering
\includegraphics[width=1.0\columnwidth]{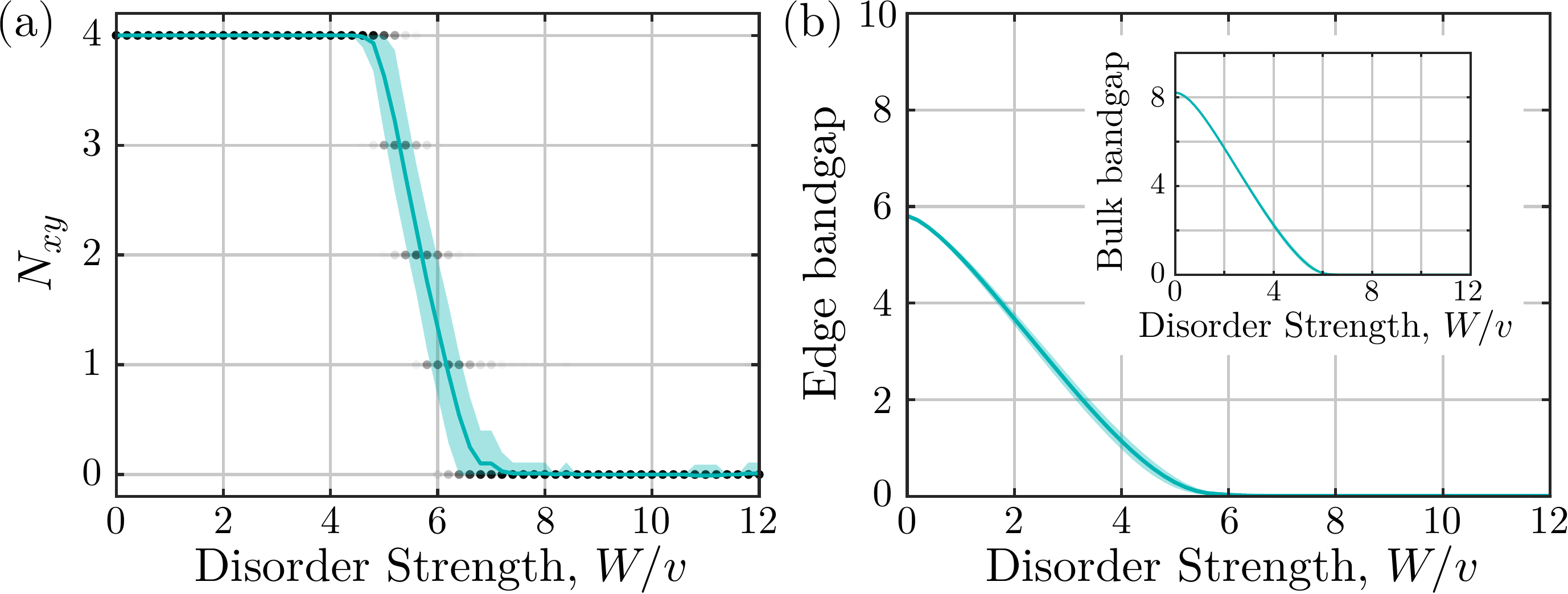}
\caption{Numerically calculated $N_{xy}$ (a), edge band gap (b), and bulk band gap (inset), as a function of disorder strength, $W/v$, for 100 independent realizations for the disorder on a $40 \times 40$ square lattice whose underlying ordered system has $w_1/v = 1$, $w_2/v = 4$, and $w_{m>2} = w_D = 0$.
The shading of the points in (a) is proportional to the number of disorder realizations that yield that invariant. The solid line and shaded region show the average of the plotted quantity and the region within 1 standard deviation of the average, respectively.}
\label{fig:rand}
\end{figure}

Even though the phases shown in Figs.~\ref{fig:1} and~\ref{fig:noc4} preserve crystalline symmetries, phases with nonzero MCNs are robust in the presence of short-range correlated disorder that breaks crystalline symmetries. To demonstrate this, we add disorder to the nearest-neighbor hopping coefficients of this model. In particular, we consider a uniform lattice with $C_{4}$ symmetry, whose disorder then breaks all spatial symmetries, as well as time-reversal symmetry, by taking values $v_{ij} \rightarrow v_{ij} + (W/\sqrt{2})(\xi_{0,ij}^{(\textrm{Re})} + \ii \xi_{0,ij}^{(\textrm{Im})})$ and $w_{1,ij} \rightarrow w_{1,ij} + (W/2\sqrt{2})(\xi_{1,ij}^{(\textrm{Re})} + \ii \xi_{1,ij}^{(\textrm{Im})})$, which for sufficiently large disorder strength $W$ causes a phase transition into a trivial phase. Here, $\xi \in [-1, 1]$ are uniformly distributed random numbers and $v_{ij}$ and $w_{1,ij}$ are the hopping strengths between neighboring lattice sites $i,j$ within the same unit cell and between adjacent unit cells, respectively. As can be seen in Fig.~\ref{fig:rand}, an $N_{xy}=4$ phase remains strictly quantized until a transition drives the system into a trivial phase with $N_{xy}=0$ when the disorder becomes sufficiently strong. This transition coincides with both bulk and edge band gap closings (up to finite size effects, see Supplemental Material~\cite{SuppInfo}).

Recently, several studies have shown that chiral symmetry alone quantizes quadrupole and octupole moments in insulators~\cite{roy2020quad,li2020quad,yang2021quad}. Our results show that protection solely due to chiral symmetry also applies to the larger family of topological phases protected by MCNs. This must be the case as systems with different MCNs also possess different numbers of topological zero-energy states at each corner; thus, to transition between them, extended zero-energy channels must exist through which some topological states delocalize and hybridize away from zero energy. Such channels are provided by bulk or boundary closings of the energy gap.

Higher-order topological phases have been found in bismuth~\cite{schindler2018bismuth} and $\mathrm{Bi_4 Br_4}$~\cite{Noguchi20}.
More recently, the mechanisms for the protection and confinement of modes of higher-order topology have found fertile ground in photonics, acoustics, and topoelectric circuits~\cite{noh2018,peterson2018,serragarcia2019,imhof2018,xue2019b,mittal2019,he2019quadrupole,bao2019,xue2019,khanikaev2019,he2020}, where they can be used to create robust cavities \cite{ota_photonic_2019,proctor_robustness_2020} and lasers \cite{zhang_low-threshold_2020,kim_multipolar_2020}. In fact, since chiral-symmetric HOTPs with large MCNs require increasingly stronger longer-range hoppings, these phases may be hard to attain in solid-state systems, where the electron's hoppings attenuate with separation. However, these phases are readily accessible in microwave photonic resonator arrays~\cite{peterson2018,kollar2019}, topoelectric circuits~\cite{imhof2018}, or sonic crystals~\cite{deng2021}, all of which can implement deformable lattice sites and couplers, which enables separating the geometric configuration of the lattice from the strength of the couplings of resonating states, thus easily achieving long-range couplings~\cite{kollar2019,deng2021}. Another candidate platform is ultracold atoms in optical lattices, where the realization synthetic gauge fields~\cite{aidelsburger2013,miyake2013,jotzu2014} and modulation of hopping terms~\cite{aidelsburger2013} in 2D has been experimentally shown. Adding long-range hoppings to this platform has been long sought after, and a recent a proposal has been put forward~\cite{martinez2021} that could give this platform access to the phases we present.

\begin{acknowledgments}
\emph{Acknowledgements.} We thank Frank Schindler, Chaoxing Liu, and Shinsei Ryu for interesting discussions. W.A.B. is thankful for the support of the Moore Postdoctoral Fellowship at Princeton University and the Eberly Postdoctoral Fellowship at the Pennsylvania State University.
A.C. acknowledges support from the Center for Integrated Nanotechnologies, an Office of Science User Facility operated for the U.S.\ Department of Energy (DOE) Office of Science, and the Laboratory Directed Research and Development program at Sandia National Laboratories. Sandia National Laboratories is a multimission laboratory managed and operated by National Technology \& Engineering Solutions of Sandia, LLC, a wholly owned subsidiary of Honeywell International, Inc., for the U.S.\ DOE’s National Nuclear Security Administration under Award DE-NA-0003525. The views expressed in the article do not necessarily represent the views of the U.S.\ DOE or the U.S. Government.
\end{acknowledgments}
\bibliography{references}

\end{document}


\title{Chiral-Symmetric Higher-Order Topological Phases of Matter:\\Supplemental Material} 

\author{Wladimir A. Benalcazar}
\email{wb7707@princeton.edu}
\affiliation{Department of Physics, Princeton University, Princeton, New Jersey 08542, USA}
\affiliation{Department of Physics, Pennsylvania State University, University Park, Pennsylvania 16802, USA}
\author{Alexander Cerjan}
\affiliation{Sandia National Laboratories, Albuquerque, New Mexico 87185, USA}
\affiliation{Center for Integrated Nanotechnologies, Sandia National Laboratories, Albuquerque, New Mexico 87185, USA}

\maketitle 

This Supplemental Material is divided into 4 sections. In Section~\ref{sec:1}, we describe the structure of chiral-symmetric Hamiltonians and how to calculate the \emph{winding number} in discrete periodic extended systems (i.e., lattice systems) in terms of a sublattice dipole moment operator, which will allow us to derive the general class of multipole chiral numbers in any dimension. In Section~\ref{sec:2}, we enunciate the generalizations of these invariants for 2D and 3D, which label topological phases with higher-order topology. We also show that these invariants are strictly quantized under chiral symmetry, and provide expressions for them in the form of Bott index. Finally, we differentiate these invariants from other first-order topological indices that protect zero-dimensional states in 2D crystals. In Section~\ref{sec:3}, we detail properties of the model proposed in the Main Text, and show the correspondence between the index $N_{xy}$ and its corner states. Finally, in Section~\ref{sec:4}, we comment on the existence of bulk and boundary obstructions separating these phases.

\section{Chiral-symmetric systems: general characteristics}
\label{sec:1}
Chiral-symmetric systems are described by Hamiltonians $\mathcal{H}$ that obey
\begin{align}
\Pi \mathcal{H} \Pi = -\mathcal{H},
\end{align}
where $\Pi$ is the \emph{chiral operator}. The degrees of freedom in chiral-symmetric systems can be divided into two sublattices, $A$ and $B$. The chiral operator is equal to
\begin{align}
\Pi=\sum_{\Rv, \alpha \in A} c^\dagger_{\Rv, \alpha} \ket{0} \bra{0} c_{\Rv, \alpha} - \sum_{\Rv, \beta \in B} c^\dagger_{\Rv, \beta} \ket{0} \bra{0} c_{\Rv, \beta}.
\end{align}
From now on, we will constrain our problem to the case in which the two sublattices have equal number of degrees of freedom, $N_A=N_B$.
For an eigenstate of the Hamiltonian $\ket{\psi_{n}}$ with energy $\epsilon_{n}$, there is a partner eigenstate $\Pi \ket{\psi_{n}}$ with opposite energy,
\begin{align}
\mathcal{H} \Pi \ket{\psi_{n}} = - \Pi \mathcal{H} \ket{\psi_{n}} = -\epsilon_{n} \Pi \ket{\psi_{n}}.
\end{align}
Thus, the energy spectrum of a chiral-symmetric system is symmetric with respect to zero energy. 

In the basis in which the degrees of freedom are ordered so that those of sublattice $A$ come first and then those of sublattice $B$, the chiral operator takes the form $\Pi = \sigma_z \otimes I_{N_A \times N_A}$, where $\sigma_z$ the third Pauli matrix and $I_{n \times n}$ is the identity matrix of size $n$. In that same basis, the Hamiltonian takes the form
\begin{align}
\mathcal{H}=\begin{pmatrix}
0 & h\\
h^\dagger & 0
\end{pmatrix}.
\label{eq:ChiralHamiltonian}
\end{align}
The eigenstates of $\mathcal{H}$ can be written as $\ket{\psi}=\frac{1}{\sqrt{2}}(\psi^A_n, \psi^B_n)^T$, where $\psi_A$ and $\psi_B$ are normalized vectors that exist only in the $A$, $B$ subspaces, respectively. The chiral partner state with opposite energy is $\Pi \ket{\psi_n}=\frac{1}{\sqrt{2}}(\psi^A_n, -\psi^B_n)^T$. Evaluating $\mathcal{H}^2\ket{\psi_n}=\epsilon_n^2 \ket{\psi_n}$ leads to the eigenvalue problems
\begin{align}
(h h^\dagger) \psi^A_n &= \epsilon_n^2 \psi^A_n \nonumber\\
(h^\dagger h) \psi^B_n &= \epsilon_n^2 \psi^B_n,
\label{eq:SVDstructure}
\end{align}
so that $\psi^A_n$ and $\psi^B_n$ can be easily obtained by diagonalizing $hh^\dagger$ or $h^\dagger h$, respectively. Equation~\eqref{eq:SVDstructure} has a structure compatible with a singular value decomposition. Indeed, it is possible to write $h$ as
\begin{align}
h=U_A \Sigma U_B^\dagger
\label{eq:SVD}
\end{align}
where $U_A$ is a unitary matrix representing the space spanned by $\{\psi^A_n\}$, i.e., $U_A=(\psi^A_1, \psi^A_2 \ldots, \psi^A_{N_A})$ and similarly for $U_B$, and $\Sigma$ is a diagonal matrix containing the singular values. Notice, for example, that
\begin{align}
hh^\dagger = U_A \Sigma^2 U_A^\dagger \nonumber \\
h^\dagger h = U_B \Sigma^2 U_B^\dagger
\end{align}
which are compatible with Eq.~\eqref{eq:SVDstructure} if we identify the squared energies $\{\epsilon^2_n\}$ with the square of the singular values in $\Sigma$.

In what follows, it will be convenient to have a `flattened' version of Hamiltonian~\eqref{eq:ChiralHamiltonian}, given by
\begin{align}
\mathcal{H}= \begin{pmatrix}
0 & q\\
q^\dagger & 0
\end{pmatrix}, \quad q = U_AU^\dagger_B,
\label{eq:HamiltonianFlattened}
\end{align}
whose energies are $\epsilon_n=\pm1$. The flattening was achieved by dropping the singular value matrix $\Sigma$ in the definition of $q$ in Eq.~\eqref{eq:HamiltonianFlattened}. Contrast this with the equivalent expression in Eq.~\eqref{eq:SVD} for Hamiltonian \eqref{eq:ChiralHamiltonian}. Although the Hamiltonians~\eqref{eq:ChiralHamiltonian} and \eqref{eq:HamiltonianFlattened} represent different systems, those two systems will belong to the same topological phase. We will see, however, that Hamiltonian \eqref{eq:HamiltonianFlattened} facilitates the calculation of the topological invariants associated with its topological phase.

\subsection{The winding number as a sublattice polarization}
We now turn our attention to chiral-symmetric Hamiltonians that are extended and periodic (i.e., lattice systems). These systems have topological phases labelled by an integer invariant, known as the \emph{winding number}. Since topological properties of a topological phase are preserved under adiabatic deformations, i.e., those deformations that preserve the energy gap and symmetry, let us consider, without loss of generality, a flattened version of Hamiltonian \eqref{eq:ChiralHamiltonian} at each crystal momentum $k$, 
\begin{align}
\mathcal{H}(k) = \begin{pmatrix}
0 & q(k)\\
q(k)^\dagger & 0
\end{pmatrix}, \quad q(k) = U_A(k)U^\dagger_B(k),
\label{eq:BlochHamiltonianFlattened}
\end{align}
where $\mathcal{H}(k)$ is a Bloch Hamiltonian with a \emph{flat} energy spectrum $\epsilon=\pm1$ at all $k$. The winding number, which classifies different homotopy classes $\pi_1[U(n)] \cong \mathbb{Z}$ in 1D, is given by
\begin{align}
N_x =\frac{\ii}{2\pi} \int_{-\pi}^\pi \tr \left[ q(k)^\dagger \partial_k q(k) \right] dk \in \mathbb{Z}.
\label{eq:WindingNumber}
\end{align}
By replacing $q(k)=U_A(k)U^\dagger_B(k)$ in Eq.~\eqref{eq:WindingNumber}, we have
\begin{align}
N_x &=\frac{\ii}{2\pi} \int_{-\pi}^\pi \tr \left[ U_A^\dagger(k) \partial_k U_A(k) \right] dk+ \frac{\ii}{2\pi} \int_{-\pi}^\pi \tr \left[ U_B(k) \partial_k U_B^\dagger(k) \right] dk.\nonumber \\
&=-\frac{1}{2\pi} \int_{-\pi}^\pi \tr \left[ \A_A(k) \right] dk + \frac{1}{2\pi} \int_{-\pi}^\pi \tr \left[ \A_B(k) \right] dk.
\label{eq:WindingPolarization}
\end{align}
Here $\A_A(k)=-\ii U_A^\dagger(k) \partial_k U_A(k)$ is the Berry connection in the $A$ sublattice, as is $\A_B$ for the $B$ sublattice. In the second step we have used $\tr[U^\dagger(k) \partial_k U(k)]= - \tr[U(k) \partial_k U^\dagger(k)]$. Notice that Eq.~\eqref{eq:WindingPolarization} is \emph{not} defined mod 1. 

\subsection{Winding number in discrete space}
We now seek to derive an expression equivalent to the winding number but in real space instead of momentum space. Such an expression would enable the characterization of disordered lattices. A first step is to discretize Eq.~\eqref{eq:WindingPolarization}. For that purpose, consider the following expression,
\begin{align}
U^\dagger(k) \partial_k U(k) \approx U^\dagger(k) (U(k+\delta k)-U(k))/\delta k, \nonumber
\end{align}
valid as $\delta k \rightarrow 0$. Rearranging this expression,
\begin{align}
I + \delta k U^\dagger (k) \partial_k U(k) \approx U^\dagger(k) U(k+\delta k). \nonumber
\end{align}
Taking the logarithm on both sides, and using the approximation $\log(1+x)\approx x$ on the LHS, we have 
\begin{align}
\delta k U^\dagger (k) \partial_k U(k) \approx \log (U^\dagger(k) U(k+\delta k)).
\end{align}
Let us now use this expression on Eq.~\eqref{eq:WindingPolarization} to have a discretized version of it,
\begin{align}
N_x &=\frac{\ii}{2\pi} \sum_k \tr \left[ \log (U_{A,k}^\dagger U_{A, k+\delta k}) - \log (U_{B,k}^\dagger U_{B, k+\delta k}) \right] \nonumber\\
&=\frac{1}{2\pi \ii} \sum_k \tr \log \left[F_{A,k} F_{B,k}^\dagger \right],
\label{eq:Invariant1DMomentumDiscrete}
\end{align}
where in the last step we used the fact that, for a unitary matrix $M$, $\log (M)=-\log(M^\dagger)$. The unitary matrices $F_{\subl,k}$ are determined by first considering the matrices
\begin{align}
 G_{\subl,k}=U_{\subl,k+\delta k}^\dagger U_{\subl,k},
 \end{align}
which are \emph{not} unitary due to the discretization of $k$~\footnote{$G_k$ becomes unitary only in the thermodynamic limit, where the spectrum $k$ is continuous.}. To restore unitatiry, consider the singular value decomposition
\begin{align}
G_{\subl,k} = V_{\subl,k} D_{\subl,k} W_{\subl,k}^\dagger,
\end{align}
where $D_{\subl,k}$ is the diagonal matrix containing the singular values. In the thermodynamic limit, $D_{\subl,k}$ is the identity matrix. For finite $L$, on the other hand, the magnitud of the diagonal elements of $D_{\subl,k}$ are less than 1. We define
\begin{align}
F_{\subl,k} = V_{\subl,k} W_{\subl,k}^\dagger
\end{align}
which is unitary.

\subsection{Winding number in position space}
Having discretized the expression for the winding number for the case in which the momenta is discrete, we now derive the position space version of the winding number~\cite{lin2021realspace}. For that purpose, consider the second-quantization operators for the anihilation of electrons in real space and crystal momentum space, which are related by the Fourier transform
\begin{align}
c_{R,\alpha} &= \frac{1}{\sqrt{L}}\sum_k e^{-\ii kR}c_{k,\alpha},\nonumber \\
c_{k,\alpha} &= \frac{1}{\sqrt{L}}\sum_R e^{\ii kR}c_{R,\alpha}.
\label{eq:FourierTransform}
\end{align}
To enforce periodic boundary conditions, we impose the constraint
\begin{align}
c_{R+L, \alpha}=c_{R,\alpha} \rightarrow k=\frac{2\pi}{L} m,\;\;m \in \mathbb{Z},
\label{eq:app_creation_operator_boundary_conditions}
\end{align}
which means that in Eq.~\eqref{eq:FourierTransform}, $k \in \delta_k(0, 1, \ldots L-1)$, where $\delta_k=2\pi/L$.

To determine the winding number, we will need to make use of the operator associated with polarization~\cite{resta1998} defined over each sublattice $\subl=A,B$~\cite{lin2021realspace},
\begin{align}
\hat{P}_x^\subl=\sum_{R,\alpha \in \subl} c^\dagger_{R,\alpha}\ket{0} \exp \left[-\ii \frac{2\pi}{L} R\right]  \bra{0} c_{R,\alpha}.
\end{align}
This operator can be written in momentum space as
\begin{align}
\hat{P}_x^\subl=\sum_{k,\alpha \in \subl} c^\dagger_{k+\delta_{k},\alpha} \ket{0} \bra{0} c_{k,\alpha}.
\end{align}
The action of $\hat{P}_x^\subl$ on a state $\ket{k,\alpha}=c^\dagger_{k,\alpha}\ket{0}$ is to shift the momentum of the state on sublattice $\subl$, i.e., $\hat{P}_x^\subl\ket{k,\alpha}=\ket{k+\delta_k,\alpha}$.

In what follows, we show that the most general expression for the winding number is
\begin{align}
\boxed{N_x =\frac{1}{2\pi \ii} \tr \log \left( \bar{P}_x^A \bar{P}_x^{B\dagger} \right)}
\label{eq:Invariant1D}
\end{align}
where $\bar{P}_x^\subl=U_\subl^\dagger P_x^\subl U_\subl$.

For that purpose, we will transform Eq.~\eqref{eq:Invariant1D} into crystal momentum space and show that it reduces to Eq.~\eqref{eq:Invariant1DMomentumDiscrete}. For that purpose, let us consider the projectors into sublattice $\subl$
\begin{align}
\mathcal{P}_\subl=\sum_{n,k,\alpha,\beta} [u^\subl_{n,k}]_\alpha c^\dagger_{k,\alpha}\ket{0} \bra{0} c_{k,\beta} [u^\subl_{n,k}]^*_\beta,
\end{align}
where $[u^\subl_{n,k}]_\alpha$ denotes the $\alpha$th component of the $n$th singular state in the SVD decomposition of Eq.~\eqref{eq:SVD}, corresponding to the $n$th column of $U_\subl$, for $\subl=A$ or $B$, and $^*$ denotes complex conjugation. Written compactly, $\mathcal{P}_\subl=U_\subl U_\subl^\dagger$. Similarly, we can spell out $q=U_AU_B^\dagger$ [Eq.~\eqref{eq:HamiltonianFlattened}] as
\begin{align}
q=\sum_{n,k,\alpha,\beta} [u^A_{n,k}]_\alpha c^\dagger_{k,\alpha}\ket{0} \bra{0} c_{k,\beta}.[u^B_{n,k}]^*_\beta \label{eq:qdef}
\end{align}
We need to calculate $\mathcal{P}_A P^A_x q P^{B\dagger}_x \mathcal{P}_B=U_A U_A^\dagger P^A_x U_A U_B^\dagger P^{B \dagger}_x U_B U_B^\dagger$. First, let us calculate
\begin{align}
\mathcal{P}_A P_x^A = \sum_{n,k,\alpha} \ket{u^A_{n,k+\delta k}}[u^A_{n,k+\delta k}]^*_\alpha \bra{0} c_{k,\alpha}\nonumber\\
P_x^{B \dagger} \mathcal{P}_B = \sum_{n,k,\alpha} c^\dagger_{k,\alpha} \ket{0}  [u^B_{n,k+\delta k}]^*_\alpha \bra{u^B_{n,k+\delta k}},
\label{eq:aux}
\end{align}
where $\ket{u^\subl_{n,k}}=\sum_\alpha [u^\subl_{n,k}]_\alpha c^\dagger_{k,\alpha} \ket{0}$. 
Using Eqs.~\ref{eq:qdef} and \ref{eq:aux}, we have
\begin{align}
\mathcal{P}_A P^A_x q P_x^{B \dagger} \mathcal{P}_B &= \sum_{k,n,m,l} \sum_{\alpha,\beta} \ket{u^A_{n,k+\delta k}}[u^A_{n,k+\delta k}]^*_\alpha [u^A_{m,k}]_\alpha[u^B_{m,k}]^*_\beta [u^B_{l,k+\delta k}]_\beta \bra{u^B_{l,k+\delta k}}\nonumber\\
&= \sum_{k,n,m,l} \ket{u^A_{n,k+\delta k}}[F_{A,k}]_{n,m} [F^\dagger_{B,k}]_{m,l} \bra{u^B_{l,k+\delta k}}\nonumber\\
&= \sum_{k,n,l} \ket{u^A_{n,k+\delta k}}[F_{A,k}F^\dagger_{B,k}]_{n,l} \bra{u^B_{l,k+\delta k}}\nonumber\\
&= \sum_{k,n,l} \ket{u^A_{n,k}}[F_{A,k-\delta k}F^\dagger_{B,k-\delta k}]_{n,l} \bra{u^B_{l,k}}.
\label{eq:aux2}
\end{align}
The expression \eqref{eq:aux2} implies that $\bar{P}_x^A\bar{P}_x^{B\dagger}=U_A^\dagger P^A_x U_A U_B^\dagger P_x^{B \dagger} U_B$ takes the form
\begin{align}
\bar{P}_x^A\bar{P}_x^{B\dagger}=
\left( \begin{array}{ccccc}
F_{A,k_0}F^\dagger_{B,k_0} & 0 & \ldots & 0\\
 0 & F_{A,k_1}F^\dagger_{B,k_1} & \ldots & 0\\
 \vdots & \vdots & \ddots & \vdots\\
 0 & 0 & \ldots & F_{A,k_{L-1}}F^\dagger_{B,k_{L-1}}\\
\end{array} \right).
\end{align}
Notice that the matrix is diagonal in $k$. Therefore, we have
\begin{align}
N_x &=\frac{1}{2\pi \ii} \tr \log \left( \bar{P}_x^A \bar{P}_x^{B\dagger} \right) \nonumber\\
&=\frac{1}{2\pi \ii} \sum_k \tr \log \left[F_{A,k} F_{B,k}^\dagger \right]
\end{align}
which is identical to the expression for the winding number~\eqref{eq:Invariant1DMomentumDiscrete}.

In contrast, the invariant that protects the quantization of the dipole moment in chiral-symmetric insulators at half-filling is
\begin{align}
    p_x=-\frac{1}{2\pi} \im \log [ \det (U_{\textrm{occ}}^\dagger P_x U_{\textrm{occ}}) \det(P_x^{\dagger1/2})],
\end{align}
where $U_{\textrm{occ}}$ is the subspace of occupied bands at half filling and $P_x=\sum_{R,\alpha} \ket{R,\alpha} \mathcal{U}_D(R) \bra{R,\alpha}$ is the position operator defined over the entire lattice~\cite{resta1998}. The dipole moment invariant is quantized to $p_x=0$ or $1/2$ under chiral symmetry, and is related to the winding number by the expression $p_x = N_x/2$ mod 1.

\section{Generalizations of the winding number to higher-order topological systems with chiral symmetry}
\label{sec:2}
As stated in the Main Text, we generalize the invariant \eqref{eq:Invariant1D} to 2D and 3D. In 2D, the invariant is
\begin{align}
\boxed{N_{xy}=\frac{1}{2\pi \ii} \tr \log \left(\bar{Q}_{xy}^A \bar{Q}_{xy}^{B\dagger} \right) \in \mathbb{Z}}
\label{eq:Invariants1}
\end{align}
where $\bar{Q}_{xy}^\subl=U_\subl^\dagger Q^\subl_{xy} U_\subl$ for $\subl=A,B$, and $Q_{xy}^\subl$ is the sublattice quadrupole moment operator
\begin{align}
Q^\subl_{xy}&=\sum_{\Rv,\alpha \in \subl} \ket{\Rv,\alpha}\text{Exp}\left(-\ii \frac{2\pi xy}{L_x L_y}\right) \bra{\Rv,\alpha}.\label{eq:Operators1}
\end{align}
Similarly, in 3D the invariant is
\begin{align}
\boxed{N_{xyz}=\frac{1}{2\pi \ii} \tr \log \left(\bar{O}_{xyz}^A \bar{O}_{xyz}^{B\dagger} \right)\in \mathbb{Z}}
\label{eq:Invariants2}
\end{align}
where $\bar{O}^\subl_{xyz}=U_\subl^\dagger O^\subl_{xyz} U_\subl$, for $\subl=A,B$, and $O_{xyz}^\subl$ is the sublattice octupole moment operator
\begin{align}
O^\subl_{xyz}&=\sum_{\Rv,\alpha \in \subl} \ket{\Rv,\alpha}\text{Exp}\left(-\ii \frac{2\pi xyz}{L_x L_y L_z}\right) \bra{\Rv,\alpha},
\label{eq:Operators2}
\end{align}

\subsection{Quantization of the real space invariants}
The invariants \eqref{eq:Invariant1D}, \eqref{eq:Invariants1}, and \eqref{eq:Invariants2} take the form
\begin{align}
N=\frac{1}{2\pi \ii} \tr \log \left(U_A^\dagger M_A U_A U_B^\dagger M^\dagger_B U_B  \right)
\label{eq:InvariantGeneralForm}
\end{align}
where $M_\subl$ (for $\subl=A,B$) is $P_x^\subl$, $Q_{xy}^\subl$, or $O_{xyz}^\subl$ in 1D, 2D, or 3D, respectively. Notice that since the matrices $M_\subl$ and $U_\subl$ are unitary, we have
\begin{align}
\det (U_A^\dagger M_A U_A U_B^\dagger M_B^\dagger U_B)= \det (M_A M^\dagger_B)=1,
\label{eq:aux3}
\end{align}
where in the last step we use the fact that $M_x^A=M_x^B$, which is true because the two sublattices have  (i) equal number of degrees of freedom in each unit cell and (ii) the same number of unit cells, from which it follows that $M_A$ and $M_B$ are identical. From \eqref{eq:aux3} it follows that tracing the logarithm of $U_A^\dagger M_A U_A U_B^\dagger M^\dagger_B U_B$ in \eqref{eq:InvariantGeneralForm} will necessarily have to give a phase that is a multiple of $2\pi i$, i.e., it will be of the form $2\pi \ii N$, with $N \in \mathbb{Z}$. This integer $N$ is indeed the topological invariant. 

\subsection{Real space invariants written as a Bott index}
Starting again with Eq.\ \ref{eq:InvariantGeneralForm}, and performing a unitary transformation using $U_B$ of the matrix product on the interior of the matrix logarithm, we can rewrite \eqref{eq:InvariantGeneralForm} as
\begin{equation}
N=\frac{1}{2\pi \ii} \tr \log \left(U_B U_A^\dagger M_A U_A U_B^\dagger M^\dagger_B U_B U_B^\dagger \right).    
\end{equation}
Noting the definition of $q = U_A U_B^\dagger$ and again that $M_A = M_B = M$ are unitary (see discussion in previous section),
\begin{align}
N &=\frac{1}{2\pi \ii} \tr \log \left(q^{-1} M_A q M_B^\dagger \right) \notag \\
&= \frac{1}{2\pi \ii} \tr \log \left(q^{-1} M q M^{-1} \right) \notag \\
&= \textrm{Bott}(q^{-1},M).
\end{align}

\subsection{Topological protection of zero-dimensional states of first-order topolgy}
Chiral-symmetric systems can also protect zero-dimensional states at topological defects. For a point defect in $d$ dimensions, $q({\bf k}, {\bf r})$ in Eq.~\eqref{eq:ChiralHamiltonian} is parametrized by $d$ momentum variables ${\bf k}$ and $d-1$ position variables ${\bf r}$, and the number of protected states is given by the homotopy class $\pi_{2d-1}[U(N)]=\mathbb{Z}$ into which $q({\bf k},{\bf r})$ falls~\cite{teo2010}, such as in the 2D Jackiw and Rossi lattice model for the protection of 0D topological states bound to vortices.
Nontrivial homotopies exist only in odd dimensional manifolds, and thus rule out, for example, the protection of 0D states at the corners of 2D crystals. In contrast, our invariants, Eqs.~(5) and (6) of the Main Text (Eq.~\eqref{eq:Invariants1} and \eqref{eq:Invariants2} here), allow for such protection, and thus are of different nature than first-order topological indices based on homotopy classes.

\section{The extended QTI model}
\label{sec:3}
In this section we describe in detail certain characteristics of the QTI model with long-range hopping terms (Eq. (7), (8), and (9) in the Main Text). We show (A) how these phases are in general outside the framework of \emph{topological quantum chemistry}~\cite{bradlyn2017,cano2018}, (B) how the phase diagram is modified when the horizontal hopping terms are different from the vertical ones, (C) the support of the corner states, (D) the correspondence between the topological index $N_{xy}$ [Eq.~(5) in the Main Text, or Eq.~\eqref{eq:Invariants1} here] and the number and chiral charge of the corner states, (E) some subtleties regarding how the phase diagrams in the Main Text were determined, and (F) the convergence of our results with disorder with system size.

\subsection{Symmetry representations of the topological phases}
The extended QTI model has the Hamiltonian of Eq.~\eqref{eq:ChiralHamiltonian} with off-diagonal term
\begin{align}
h({\bf k}) = h_{QTI}({\bf k})+C_s h_{\textrm{SLR}}({\bf k})+ C_d h_{\textrm{DLR}}({\bf k})
\label{eq:extendedQTI}
\end{align}
where $h_{\textrm{QTI}}({\bf k})$, $h_{\textrm{SLR}}({\bf k})$, and $h_{\textrm{DLR}}({\bf k})$ are defined in Eqs.~(7), (8), and (9) of the Main Text. 

Consider the case $C_s=1$, $C_d=0$. In that case, the Hamiltonian is $C_{4v}$ symmetric and supports bulk-obstructed topological phases. At high-symmetry points (HSPs) and lines of the BZ, the representations of the elements of $C_{4v}$ are given in Table~\ref{tab:QTIirreps}. To capture more directly the topological protection due to crystalline symmetries, Table~\ref{tab:QTIirreps} also shows the symmetry indicator invariants, defined as
\begin{align}
[M_j] = \# M_j - \#\Gamma_j \label{eq:SymmetryIndicators}
\end{align}
where  $\# M_j$ is the number of bands below the gap with $C_4$ symmetry representations $M_j=e^{\ii \pi (2j-1)/4}$, for $j=1,2,3,4$, at the ${\bf M}$ point of the BZ, and similarly for the ${\bf \Gamma}$ point.

\begin{table}[h]
\begin{tabular}{c|cc|c}
\hline
\hline
phase & irreps at ${\bf \Gamma}$ & irreps at ${\bf M}$ & ($[M_1]$, $[M_2]$, $[M_3]$, $[M_4]$) \\
\hline
\hline
$N_{xy}=0$ & $\{e^{\ii 3\pi/4},e^{-\ii 3\pi/4}\}$ & $\{e^{\ii 3\pi/4},e^{-\ii 3\pi/4}\}$ & (0,0,0,0)\\
$N_{xy}=1$ & $\{e^{\ii 3\pi/4},e^{-\ii 3\pi/4}\}$ & $\{e^{\ii \pi/4},e^{-\ii \pi/4}\}$ & (1,-1,-1,1)\\
$N_{xy}=4$ & $\{e^{\ii 3\pi/4},e^{-\ii 3\pi/4}\}$ & $\{e^{\ii 3\pi/4},e^{-\ii 3\pi/4}\}$ & (0,0,0,0)\\
\hline
\hline
\end{tabular}
\caption{
$C_4$ symmetry representations and symmetry indicator invariants \eqref{eq:SymmetryIndicators} for the lowest two bands of Hamiltonian \eqref{eq:extendedQTI} at the $C_4$ invariant points of the BZ ${\bf \Gamma}$ and ${\bf M}$ for all topological phases. At all $C_2$-invariant points, the representations are $\pm \ii$. Similarly, at all $M_x$ and $M_y$ invariant lines, the representations are $\pm 1$. In both cases, those representations lead to trivial symmetry indicator invariants for $C_2$ and reflection symmetries.}
\label{tab:QTIirreps}
\end{table}

A result worth noting from Table~\ref{tab:QTIirreps} is that, while the $N_{xy}=1$ phase (i.e., the QTI nontrivial phase with $q_{xy}=e/2$) has nontrivial symmetry indicator invariants, the $N_{xy}=4$ does not. Consequently, the topology of this last phase cannot be captured by the theory of induction of representations and symmetry indicator invariants~\cite{bradlyn2017,cano2018}. Nevertheless, its nontrivial nature is evident in both the existence of a nonvanishing bulk invariant $N_{xy}=4$ and the existence of an equal number of topological zero-energy states at each corner [Fig.~1(d) of the Main Text and Fig.~\ref{fig:states1} here].

The representations due to $C_2$ symmetry and reflection symmetries are equal at all HSPs of the BZ in all phases, and thus have trivial symmetry indicator invariants.

\subsection{Phase diagram for the ordered $C_{2v}$ symmetric system}

\begin{figure}[h!]
\centering
\includegraphics[width=0.45\columnwidth]{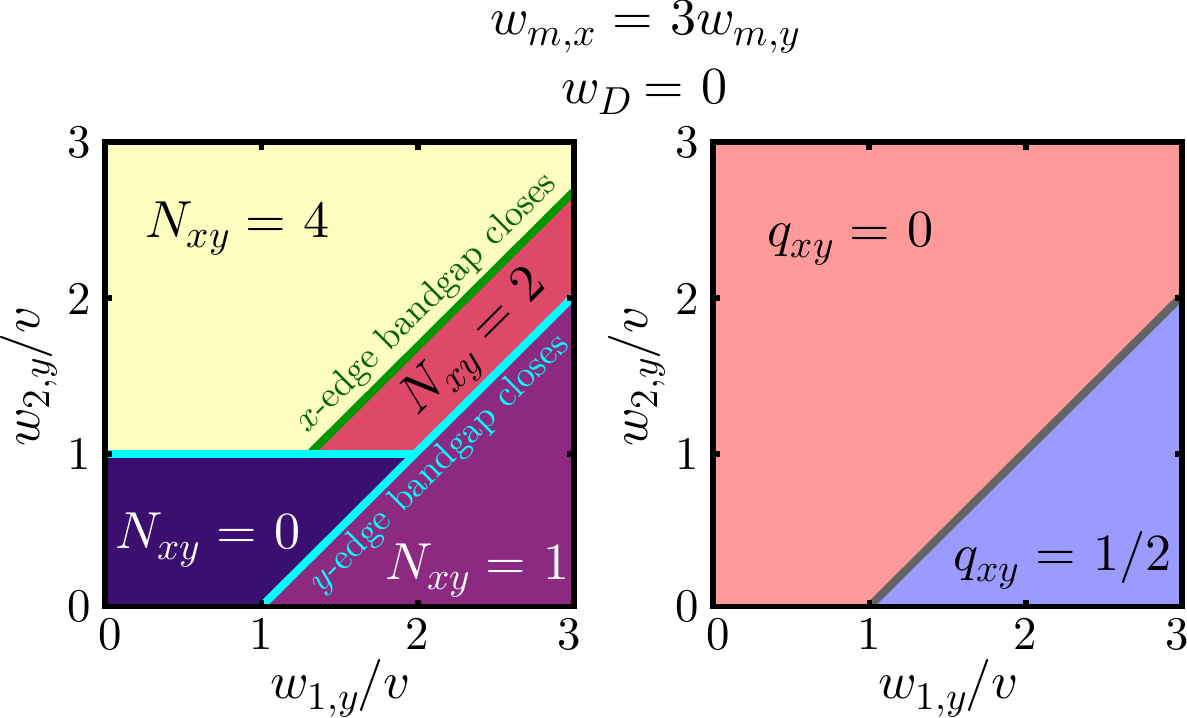}
\caption{Phase diagram of the chiral-symmetric second-order topological phase, $N_{xy}$ (left), and the quadrupole phase $q_{xy}$ (right), for a $C_{2v}$ symmetric system with open boundaries as a function of the hopping ratios, $w_{1,y}/v$ and $w_{2,y}/v$, with $w_{m,x} = 3 w_{m,y}$ and $w_D = 0$. Phase transitions where the boundary band gap closes along the $x$ edge are shown in green, while those which close along the $y$ edge are shown in cyan.}
\label{fig:c2v}
\end{figure}

In the interest of completeness, in Fig.~\ref{fig:c2v} we show the phase diagram for a system which is $C_{2v}$ symmetric, with $w_{m,x} = 3w_{m,y}$, $w_D = 0$, and $w_{m\ge 0} = 0$.

\subsection{Corner states and their support}
\begin{figure}[h!]
\centering
\includegraphics[width=.95\columnwidth]{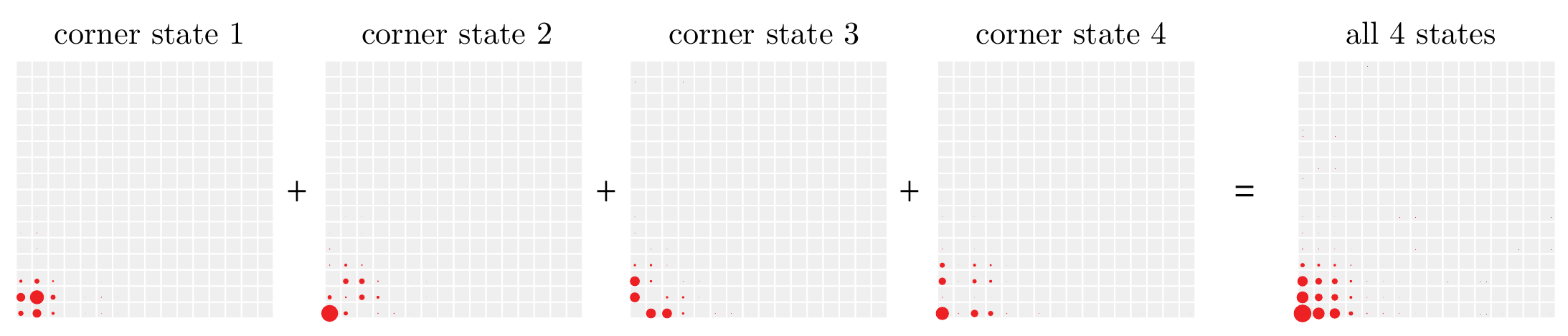}
\caption{Probability density functions of the four corner states at one of the corners of a $N_{xy}=4$ phase (four left plots) and cumulative PDF for all the four states (rightmost plot). Support at each sublattice is indicated by blue and red colors, respectively. All corner states in the same corner have the same chiral charge and thus have support on only one sublattice. In this simulation, $v=1$, $w_1=2$, $w_2=2$, $w_D=0$.}
\label{fig:states1}
\centering
\includegraphics[width=.40\columnwidth]{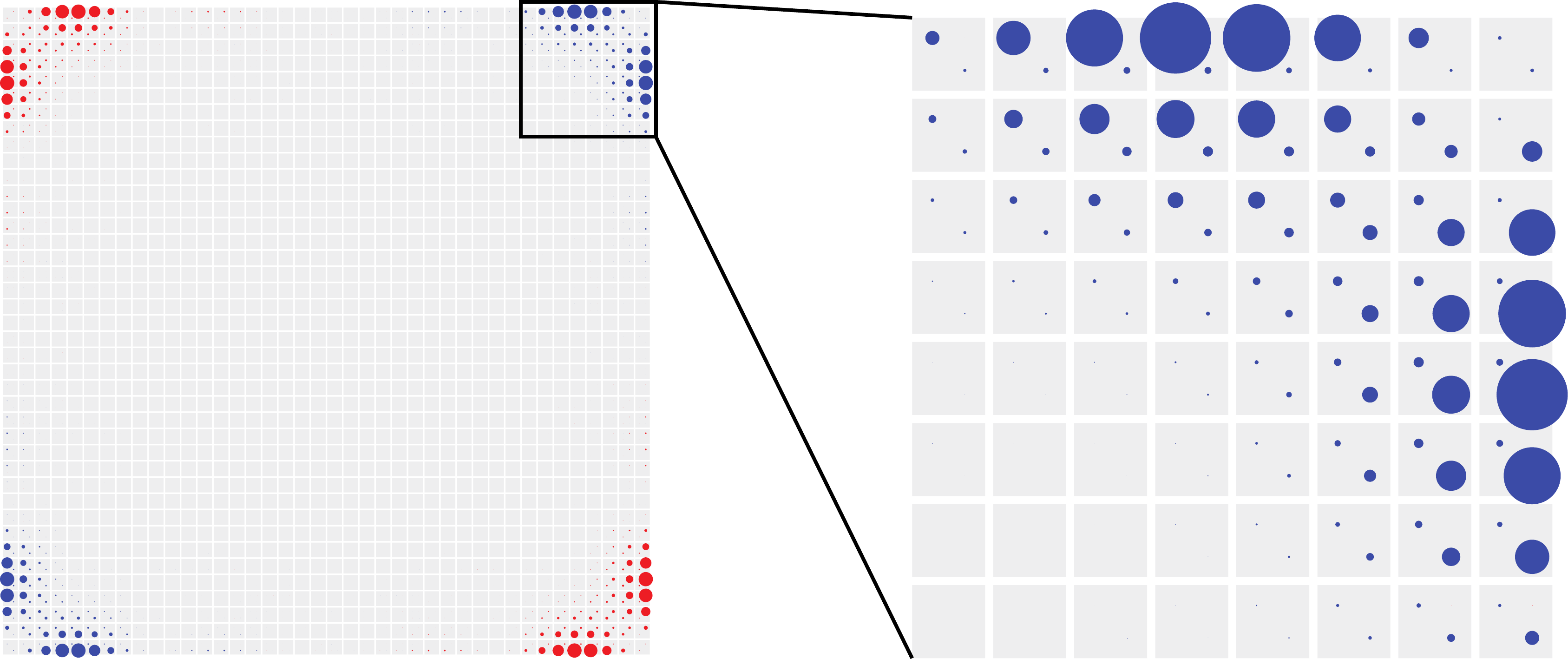}
\caption{Probability density functions of all the corner states of a $N_{xy}=-1$ phase (4 states in total, one per corner). Support at each sublattice is indicated by blue and red colors, respectively. In this simulation, $v=1$, $w_1=1$, $w_2=0.8$, $w_D=0.5$.}
\label{fig:states2}
\end{figure}

In this section, we show the form and localization of two sets of corner states for phases indicated in the Main Text. First, in Fig.~\ref{fig:states1}, we show one possible choice of the four corner-localized states of the lattice considered in Fig.~1 in the Main Text with $N_{xy} = 4$. As can be seen, the summation of the probability density functions of the four orthogonal states equals that of the total set of four states in a single corner, as shown in the Main Text. Second, in Fig.~\ref{fig:states2} we show the probability density function of the corner states of a $N_{xy}=-1$ phase from the system considered in Fig.~2 in the Main Text. Notice that the corner states have support on the opposite sublattice as those in the $N_{xy}=4$ phase (Fig.~\ref{fig:states1}) due to the reversal in the sign of the bulk index $N_{xy}$.

\subsection{Correspondence between $N_{xy}$, band gap closings, and corner states}
To claim that the multipole chiral numbers introduced here are the topological invariant connected with the appearance of corner-localized modes across a band gap closing, one must show that all three phenomena are causally connected. In other words, changes in either $N_{xy}$ or the number (or type) of corner-localized modes necessarily imply a change in other quantity, and moreover that these changes can only occur at a band gap closing.
\begin{figure}[h!]
\centering
\includegraphics[width=0.85\columnwidth]{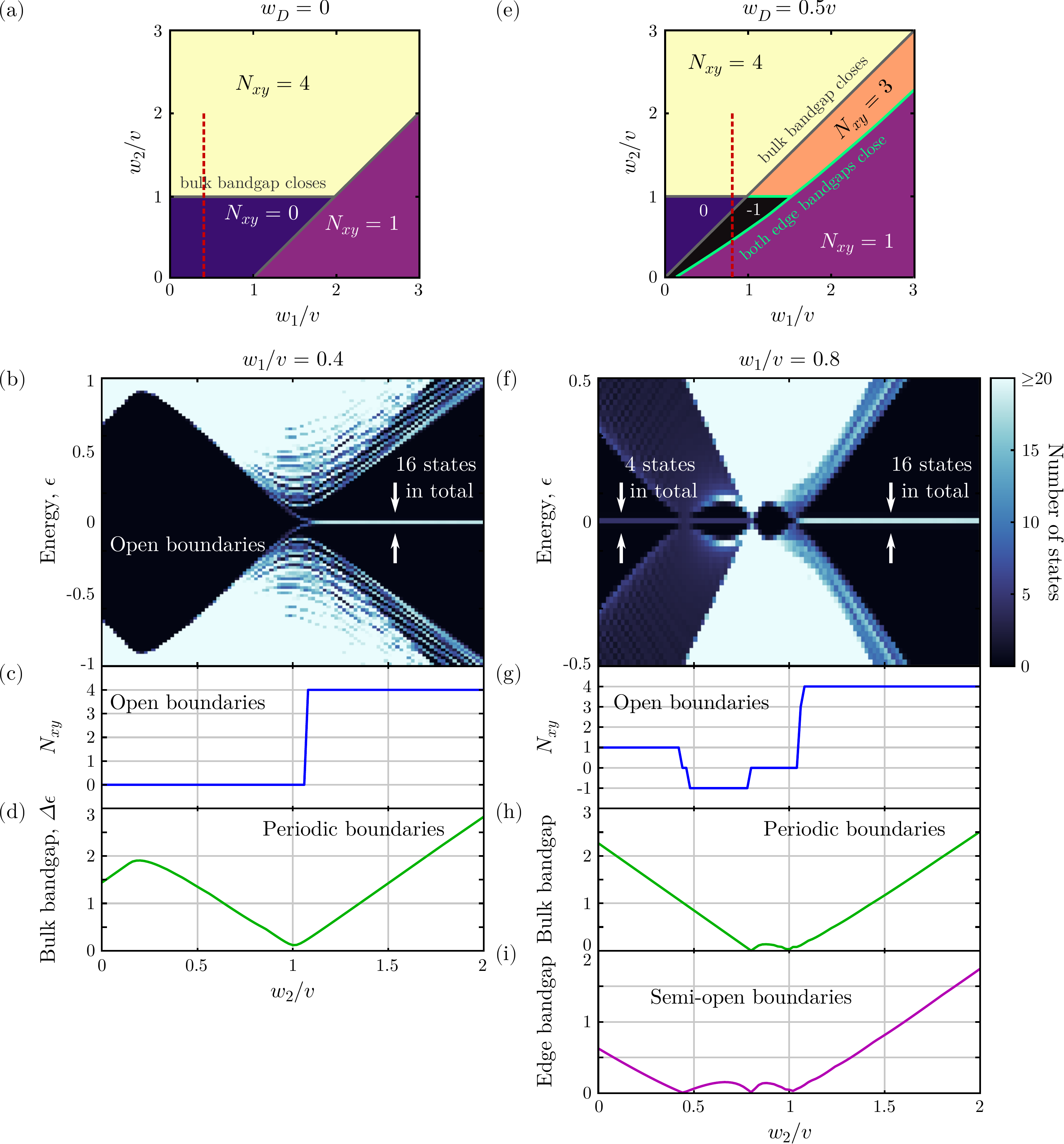}
\caption{Correspondence between the appearance of zero energy states, change in the topological invariant $N_{xy}$, and the closing of the bulk and edge band gaps for two of the systems considered in the Main Text in Figs.~1(b) and 2(a). (a) Copy of Fig.~1(b) from the Main Text, red line indicates the choice of $w_{1}/v$ and $w_{2}/v$ in (b-d). (b) Number of states relative to their energy and $w_{2}/v$ calculated with open boundary conditions. (c) $N_{xy}$ as a function of $w_{2}/v$ calculated using open boundary conditions. (d) Bulk band gap, $\Delta \epsilon$ as a function of $w_{2}/v$ calculated using periodic boundary conditions. (e-h) Similar to (a-d), except for the system in Fig.~2(a) of the Main Text. (i) Edge band gap calculated using a semi-open boundary, i.e.\ periodic in $x$, but open in $y$.
In all simulations of both systems, the system size was $80 \times 80$ unit cells.}
\label{fig:corr}
\end{figure}

To support this claim, Fig.~\ref{fig:corr}(a-d) presents numerical simulations of all three of these quantities, $N_{xy}$, the distribution of states in energy, and the bulk band gap, for the long-range quadrupole topological insulator model considered in Fig.~1(b) of the Main Text with fixed $w_1/v = 0.4$. As can be seen, the closing of the bulk band gap at $w_2/v = 1$ coincides with 8 states departing from each of the upper and lower bulk bands and becoming pinned at $\epsilon = 0$. These 16 states in total are those corner-localized states shown in Fig.~1(d) of the Main Text, which are predicted to exist as $N_{xy} = 4$ for $w_2/v > 1$. The slight discrepancy between the change in $N_{xy}$ and the closing of the bulk band gap is due to finite system effects, where the topological invariant does not change until the bulk band gap is larger that the approximate coupling strength between the corner states in adjacent corners.

In Fig.~\ref{fig:corr}(e-i), we show this same correspondence for the quadrupole topological insulator with both long-range straight and long-range diagonal hoppings considered in Fig.~2(b) of the Main Text with fixed $w_1/v = 0.8$. As $w_2/v$ is increased from $0$ to $2$, this system undergoes three separate phase transitions. The first transition, $N_{xy} = 1 \rightarrow -1$ near $w_{2}/v = 0.4$, occurs at a closing of the edge band gap, not the bulk band gap. This particular phase transition also does not change the number of corner-localized states, as there is only one state localized to each corner in both phases. Rather, across the phase transition, the corner states swap the sublattices on which they are supported. The other two phase transitions, $N_{xy} = -1 \rightarrow 0$ and $N_{xy} = 0 \rightarrow 4$, both occur when the bulk band gap closes, and result in changes in the number of corner-localized states. Again, discrepancies between the change in $N_{xy}$ from the band gap closings are due to finite system size effects.

\subsection{Determination of the phase diagrams of the extended QTI model}
There are some subtleties in how the phase diagrams shown in Figs.~1(b) and 2(a) of the Main Text are constructed due to finite system size effects. As was shown in Fig.~\ref{fig:corr}, computing the topological invariants $N_{xy}$ in finite systems causes the topological invariant not to change at exactly the same location where the band gap closes. Moreover, it is not numerically feasible to calculate an entire 2D phase diagram at a reasonable sampling density even at the system sizes shown in Fig.~\ref{fig:corr} ($80 \times 80$ unit cells), due to the memory requirements necessary to calculate the full singular value decomposition of $H$. In the definition of $N_{xy}$, every single singular value is necessary to achieve accurate results.

\begin{figure}[h!]
\centering
\includegraphics[width=0.50\columnwidth]{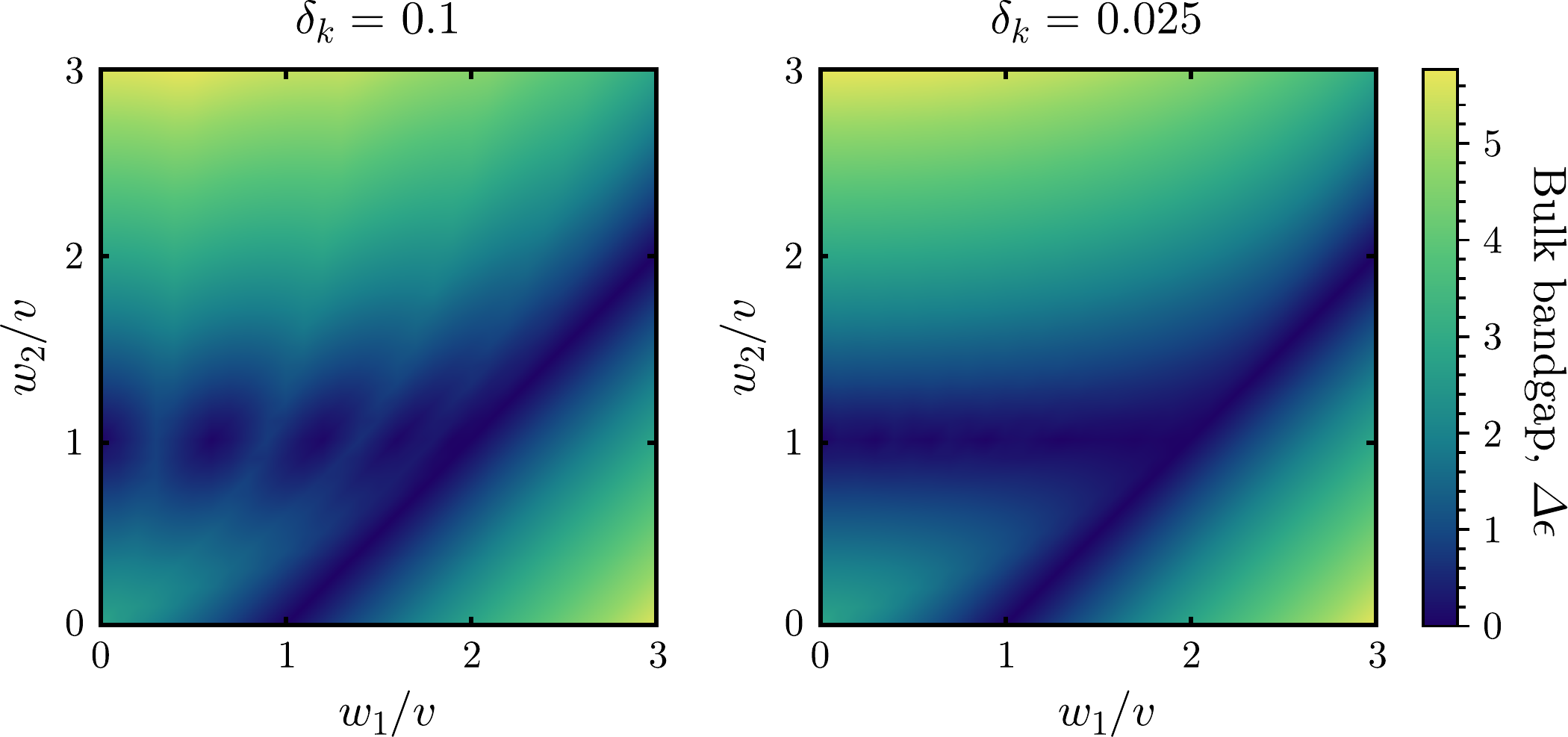}
\caption{Bulk band gaps of the $C_{4v}$ symmetric system considered in Fig.~1(b) of the Main Text calculated using periodic boundary conditions with the spacing between adjacent points in the Brillouin zone being $\delta k = 0.1$ (left) and $\delta k = 0.025$ (right).}
\label{fig:dk}
\end{figure}

As such, the phase diagrams shown in Figs.~1b, 2a, and 2b are constructed by first finding closings in the bulk and edge band structures, assuming the system is infinite in any non-open direction. Numerically, artifacts can appear in the calculation of the band gap closings due to finite system sizes, as shown in Fig.~\ref{fig:dk}, but simulations strongly suggest that the phase diagrams converge to those shown in the Main Text in the thermodynamic limit. Then, we calculate $N_{xy}$ within each region at an ensemble of points using open boundaries in all directions and using large system sizes. We also verify these calculations through direct calculation of $N_{xy}$ across the entire 2D parameter space but using smaller system sizes. 

\subsection{Finite system size effects in disordered systems}
\begin{figure}[h!]
\centering
\includegraphics[width=0.35\columnwidth]{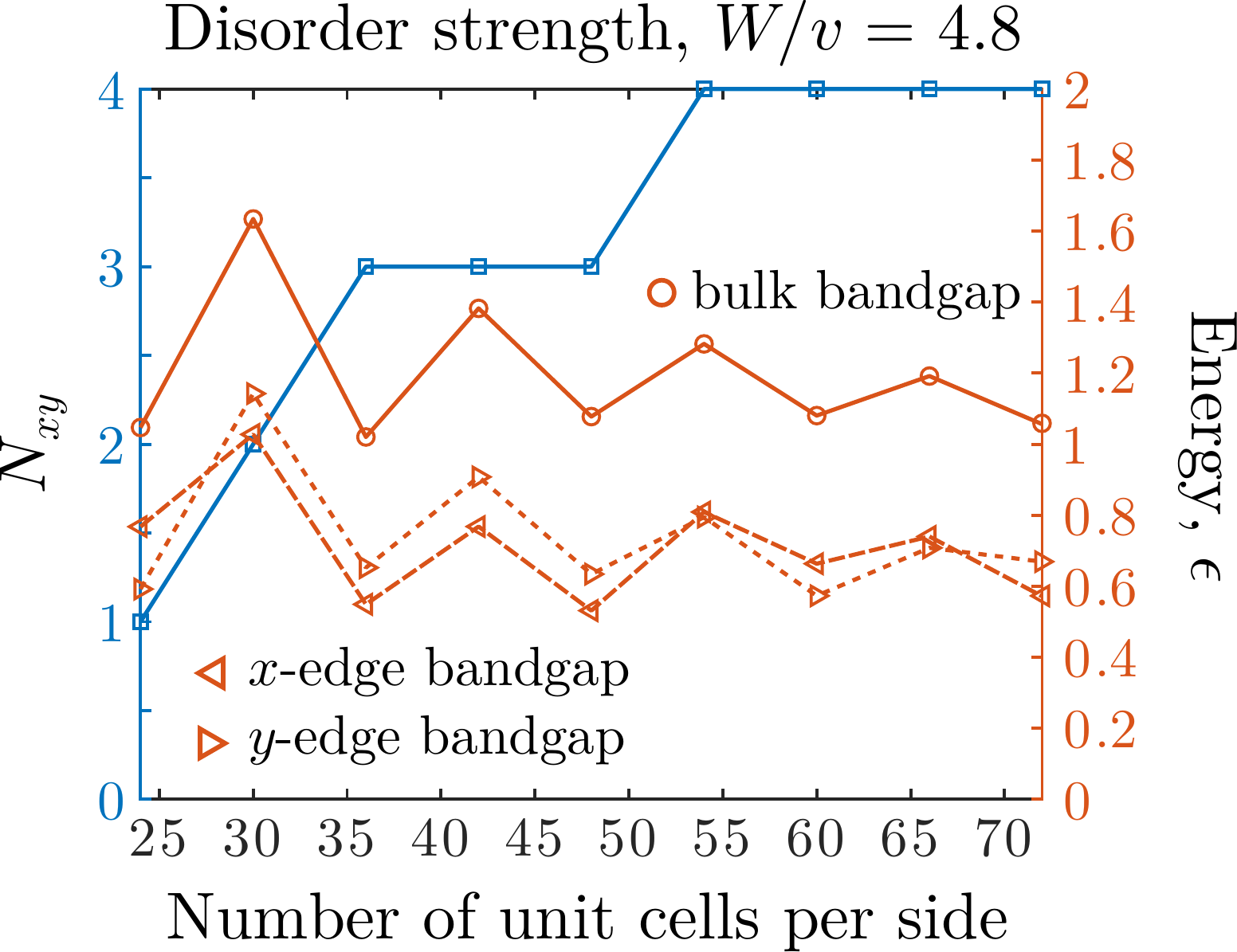}
\caption{Dependence of the topological invariant $N_{xy}$, the bulk band gap, and the two edge band gaps on the number of unit cells per side in a square lattice. The underlying system has $C_{4v}$ symmetry, while the added disorder breaks all of the symmetries (including $C_{4}$) except for chiral symmetry.}
\label{fig:nucDis}
\end{figure}

Another manifestation of the difficulties associated with finite system sizes is in calculating the topological invariant and band gaps for disordered systems. In Fig.~\ref{fig:nucDis} we show the variation and (slow) convergence of both the topological invariant, $N_{xy}$, as well as the bulk and edge band gaps as a function of system size. Here, the system contains some disorder with a fixed strength, $W/v = 4.8$, with $W$ defined in the Main Text. The underlying system possess $C_{4v}$ symmetry, with $W_1/v = 1$ and $w_2/v = 4$, while the added disorder breaks all of the system's symmetries (including $C_4$) except for chiral symmetry. Only a single realization of the disorder is shown here at each system size.

\section{Bulk and boundary energy gap closing at phase transitions}
\label{sec:4}
As shown in the Main Text, the phase transitions between phases of different multipole chiral numbers close the gap at either the boundary or the bulk. This has to be the case since changing the winding number across the phase transition goes hand in hand with the delocalization of the topological corner states, which necesitates extended gapless channels. The existence of these gapless extended channels is necessary because the hybridization of topological states away from zero energy can only occur by fusing pairs of topological states having opposite chiral charge, and such pairs do not exist within any single corner. 

\begin{figure}[h!]
\centering
\includegraphics[width=0.50\columnwidth]{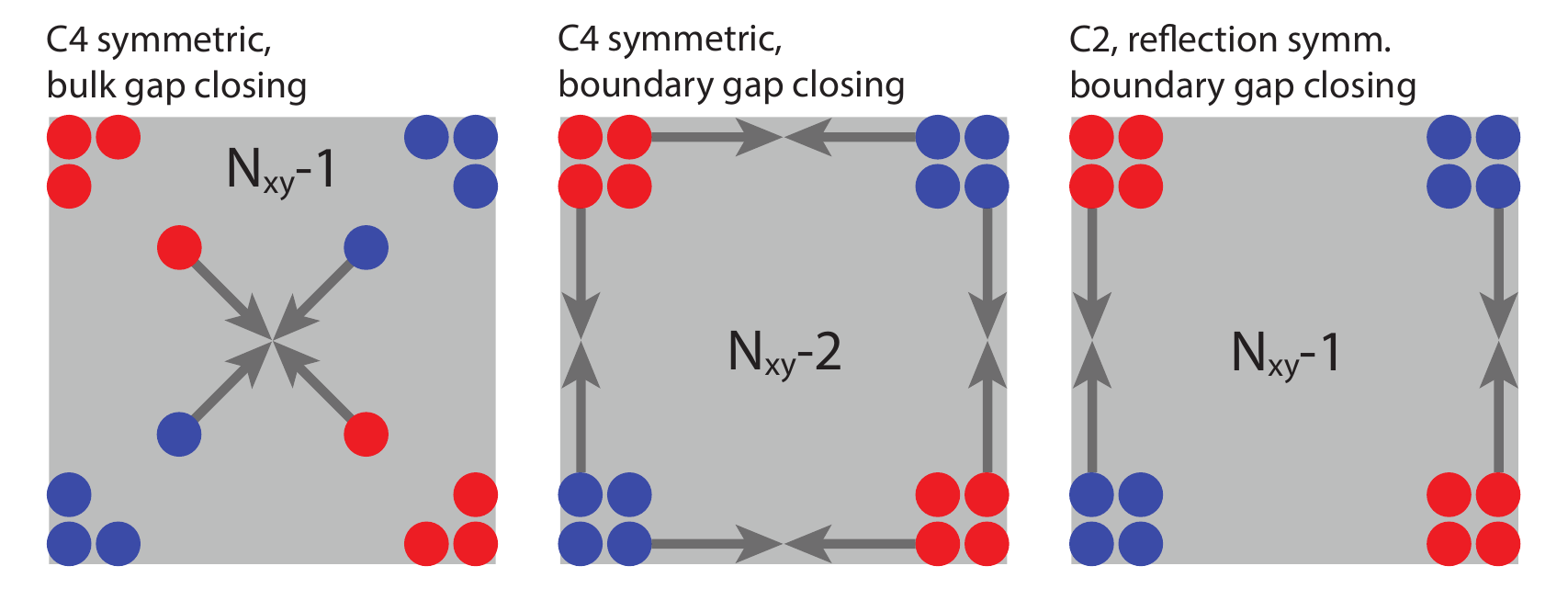}
\caption{Schematics that illustrates the hybridization of topological corner states during phase transitions. Red and blue dots represent topological corner states of opposite chiral charge. Topological corner states hybridize only if they have opposite chiral charge and bulk (a) or edge (b,c) low energy channels are available for their delocalization.}
\label{fig:botp1}
\end{figure}

This is illustrated in  Fig.~\ref{fig:botp1}. At the phase transition critical points, the bulk (a) or edges (b,c) close the energy gap, allowing the topological corner states to delocalize and hybridize away from zero energy. In (a), a $C_4$ symmetric lattice within a bulk-obstructed phase transition hybridizes corner states via bulk low energy channels, reducing the bulk invariant by 1. In (b), a $C_4$ symmetric boundary-obstructed phase transition hybridize the corner states via the boundary, reducing the bulk invariant by 2. In (c), a $C_2$ symmetric boundary obstructed transition reduces the bulk invariant by 1. 

\bibliography{supplementRefs}